\documentclass[prb,aps,twocolumn,superscriptaddress,longbibliography]{revtex4-2}
\usepackage{graphicx,color}
\usepackage{amsthm}
\usepackage{amsfonts}
\usepackage{algorithmic}
\usepackage{enumerate}
\usepackage{latexsym}
\usepackage{amsmath}
\usepackage{amssymb}
\usepackage{multirow}
\usepackage{array}
\usepackage{diagbox}
\DeclareUnicodeCharacter{2212}{\ensuremath{-}}
\usepackage{subfigure}
\usepackage[colorlinks=true,citecolor=blue,linkcolor=blue]{hyperref}

\def\avg#1{\left\langle#1\right\rangle}

\emergencystretch=\maxdimen
\begin{document}
\title{Disorder-dependent superconducting pairing symmetry in doped graphene}
\author{Kaiyi Guo$^{*}$}
\affiliation{School of Physics and Astronomy, Beijing Normal University, Beijing 100875, China\\}
\author{Yue Zhang}
\email{These authors contributed equally to this work.}
\affiliation{School of Physics and Astronomy, Beijing Normal University, Beijing 100875, China\\}
\author{Ying Liang}
\affiliation{School of Physics and Astronomy, Beijing Normal University, Beijing 100875, China\\}
\affiliation{Key Laboratory of Multiscale Spin Physics(Ministry of Education), Beijing Normal University, Beijing 100875, China\\}
\author{Tianxing Ma}
\email{txma@bnu.edu.cn}
\affiliation{School of Physics and Astronomy, Beijing Normal University, Beijing 100875, China\\}
\affiliation{Key Laboratory of Multiscale Spin Physics(Ministry of Education), Beijing Normal University, Beijing 100875, China\\}

\begin{abstract}
Disorder and doping have profound effects on the intrinsic physical mechanisms of superconductivity. In this paper, we employed the determinant quantum Monte Carlo method to investigate the symmetry-allowed superconducting orders on the two-dimensional honeycomb lattice within the Hubbard model, using doped graphene as the carrier, focusing their response to bond disorder. Specifically, we calculated the pairing susceptibility and effective pairing interactions for the $d+id$ wave and extended $s$-wave pairings for different electron densities and disorder strengths. Our calculations show that at high electron densities, increased disorder strength may lead to a transform from $d+id$ wave dominance to extended $s$ wave dominance. However, at lower electron densities, neither of the two superconducting pairings appears under larger disorder strength. Our calculations may contribute to a further understanding of the superconducting behavior in doped materials affected by disorder.
\end{abstract}

\maketitle

\noindent
\section{Introduction}

Unconventional superconductivity is one of the most important subjects in modern condensed matter physics, and in past decades, the fabrication of superconducting materials and understanding on the pairing mechanism have attracted the attention of numerous researchers  \cite{RevModPhys.63.239,PhysRevLett.43.1892}.
Inevitably, the experimental preparation of superconducting materials with doping electrons or holes shall introduce disorder into the system,
and the presence of disorder will alter the material's properties  \cite{PhysRevB.106.085429}. For example, researchers have found that disorder may raise  \cite{doi:10.1073/pnas.1817134116, PhysRevB.103.155133} or decrease  \cite{PhysRevB.106.224511, PhysRevB.79.064507} the superconducting transition temperature, or may have no effect on it  \cite{Chakraborty2022, PhysRevB.79.064507}.
In comparison to clean systems, stronger disorder increases spatial inhomogeneity, thereby enhancing localization and the superconducting energy gap  \cite{PhysRevB.68.180503}. Through numerical computations, the researchers have found that disorder leads to the formation of islands with higher superconducting order, consequently causing the system to undergo a superconducting-insulator transition   \cite{PhysRevB.108.235142,Dubi2007,Wang2021,Cao2021}. However, it is worth noting that correlated disorder may have a contrasting influence on the system, rendering superconductivity more robust  \cite{Neverov2022}. Related experimental results also respond to the complicated response of superconductivity to disorder  \cite{PhysRevB.108.134511,Kamppinen2023,Arean2015}.

Behind all of the above marvelous phenomena, the role played by disorder remains a question of great interest. To investigate the influence of disorder on superconductivity, an essential idea is to explore its effect on the superconducting pairing symmetry.
As one of the key superconducting order parameters, the pairing symmetry plays an important role in the study of superconductivity, determining the physical properties of the superconducting state  \cite{doi:10.1143/JPSJ.81.011013,PhysRevB.108.144508}, which attract the attention of both experimental and theoretical researchers  \cite{Benhabib2021, Cheng2024,PhysRevB.79.052502}. After extensive research since the discovery of doped cuprates, the pairing mechanism of superconductivity continues to remain enigmatic. It has become relatively clear that $d$-wave pairing dominates in doped cuprates  \cite{RevModPhys.72.969,Bednorz1986,PhysRevLett.73.593}, while iron-based superconductors are primarily driven by $s_{\pm}$ wave pairing  \cite{PhysRevLett.101.057003,PhysRevB.78.134512}. The situation becomes more intricate in the case of nickel-based superconductors, in which both
$d$-wave and $s$-wave superconducting gaps are observed  \cite{PhysRevResearch.1.032046,PhysRevB.106.195112,PhysRevB.102.220501}.

Recently, superconductivity in graphene-based systems has attracted a lot of attention \cite{PhysRevB.108.134514,DiBernardo2017,Perconte2018,Cao2018Correlated, Cao2018Unconventional,PhysRevB.101.155413, PhysRevB.98.241407,HUANG2019310,PhysRevB.99.121407,PhysRevB.98.235158,PhysRevB.108.134515,PhysRevB.90.094516,PhysRevB.104.035104,PhysRevB.106.104506,
PhysRevB.81.085431,PhysRevB.80.094522,PhysRevB.84.121410,PhysRevB.93.241410,PhysRevLett.127.217001}. Researchers have shown evidence for triggered superconducting density of states by placing monolayer graphene on $d$-wave superconductors \cite{DiBernardo2017}. Subsequently, studies have also demonstrated gate-tunable high-temperature superconducting proximity effects in monolayer graphene \cite{Perconte2018}. Since march of 2018, the observation of superconducting phenomena on magic-angle graphene superlattices generated great excitement \cite{Cao2018Correlated, Cao2018Unconventional}. Among them, different pairing symmetry, for example, $d+id$ \cite{PhysRevB.101.155413, PhysRevB.98.241407,HUANG2019310}, $p+ip$ \cite{PhysRevB.99.121407,PhysRevB.98.235158}, as well as extensive $s$ ($ES$)  \cite{PhysRevB.108.134515}waves are proposed in different situations or within different models. Thus, the doped graphene-based materials provide an interesting platform for investigating superconducting pairing symmetries, which may lead further understanding on the pairing mechnism \cite{PhysRevB.90.094516,PhysRevB.104.035104,HUANG2019310,PhysRevB.106.104506}.

\begin{figure}[htbp]
\centering
\includegraphics[scale=0.45]{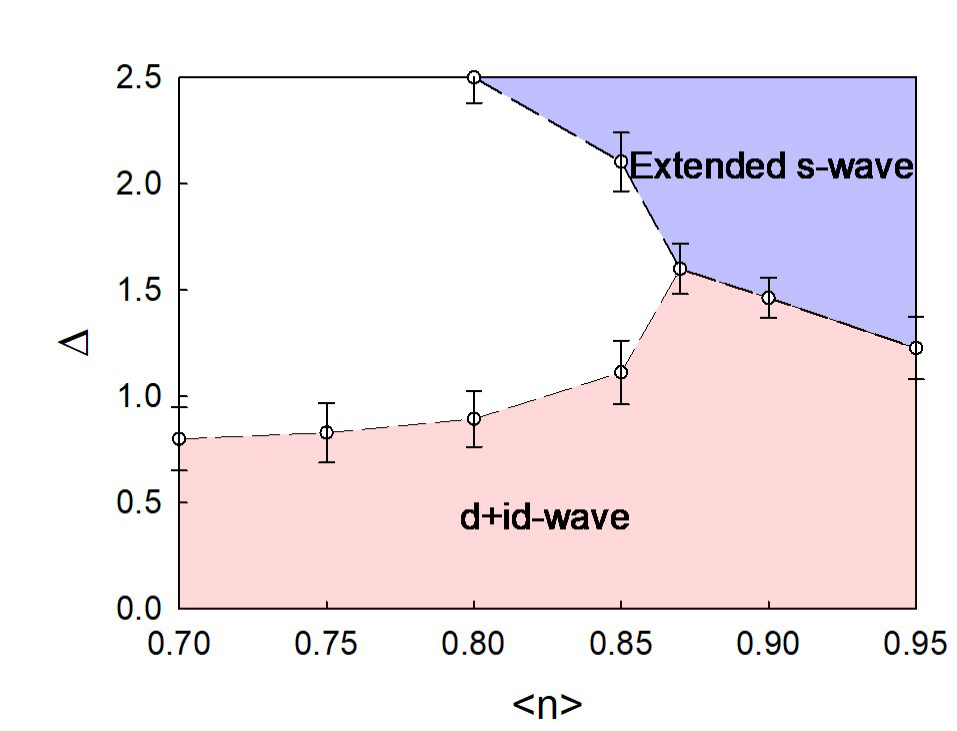}
\caption{Illustration of superconduction pairing of doping-dependent disordered Hubbard model on honeycomb lattice, considering the lattice size of $N_s=96$, and the temperature of $\beta=10$. $\Delta$ labels the disorder strength and $\avg{n}$ represents the electron density.}
\label{Fig:phase}
\end{figure}

Naturally, doping and disorder are both present in real materials \cite{PhysRevLett.103.056404,Osofsky2016}. In this paper we investigated the dominant superconducting pairing symmetry in doped graphene with bond disorder. The disorder induced a finite density of state in lightly doped graphene, which is different from clean system. It is found that the introduced disorder not only increase the effective pairing interaction, but also lead to a transform in the dominant pairing symmetry. Our primary findings are presented in the form of a superconducting pairing in Fig.\ref{Fig:phase}, which shows that the symmetry of pairing is not only related to the strength of disorder, but also to the electron density. When the strength of disorder is small, pairing with $d+id$ symmetry dominates with $ES$ symmetry, which is consistent with previous results \cite{PhysRevB.84.121410}. However, the situation is different when the disorder strength is large. Near half-filling where the electron density is large, with the increasing of the disorder strength, the $d+id$ wave is suppressed, while the $ES$ wave is strengthened, and when the disorder strength exceeds a critical value, the system's superconducting pairing shall be dominated by the $ES$ wave. Through the analysis of the effective pairing interaction $P_{\alpha}-\widetilde{P}_{\alpha}$, we find that in instances of large disorder strength and low electron density, both value of $P_{d+id}-\widetilde{P}_{d+id}$ and $P_{ES}-\widetilde{P}_{ES}$ are negative and there is no longer an effective attractive interactions between the electrons. Consequently, neither $d+id$ wave nor the $ES$ wave exists. The actual situation in this case may be even more intricate. More detailed computational results will be discussed in the following sections.

\noindent

\section{Model and methods}

The low-energy electronic and superconducting property of graphene can be well described by the disordered Hubbard model on a honeycomb lattice \cite{RevModPhys.81.109}:
\begin{eqnarray}
\hat H=&&-\sum_{{\bf \avg{i,j}},\sigma}t_{\bf ij}(\hat c_{{\bf i}\sigma}^\dagger \hat c_{{\bf j} \sigma}^{\phantom{\dagger}}+\hat c_{{\bf j}\sigma}^\dagger \hat c_{{\bf i} \sigma}^{\phantom{\dagger}})-\mu \sum_{{\bf i}\sigma} \hat n_{{\bf i}\sigma}\notag\\
&&+U\sum_{{\bf i}}\hat n_{{\bf i}\uparrow}\hat n_{{\bf i}\downarrow}
\label{hamiltonian}
\end{eqnarray}
Here, $\hat c_{{\bf i}\sigma}^\dagger(\hat c_{{\bf j} \sigma}^{\phantom{\dagger}})$ is the creation (annihilation) operator of an electron with spin $\sigma$ at site ${\bf i}({\bf j})$, and $\hat n_{{\bf i}\sigma}=\hat c_{{\bf i}\sigma}^\dagger\hat c_{{\bf j} \sigma}^{\phantom{\dagger}}$ is the number operator, denoting the number of spin-$\sigma$ electrons at site ${\bf i}$. $\mu$ is the chemical potential which determines the density of the system, when $\mu=\frac{U}{2}$, $\avg{n}=1$, the system is half-filled, indicating the particle-hole symmetry. When deviating from half-filling, the corresponding $\mu$ for a specific $\avg{n}$ varies with a different set of parameters. Therefore, for each set of parameters, we have individually tuned the chemical potential $\mu$ to fix the specific electronic density $\avg{n}$. $t_{\bf ij}$ represent the hopping amplitude between two nearest-neighbor sites ${\bf i}$ and ${\bf  j}$, and the bond disorder is induced by modifying the matrix element $t_{\bf ij}$ of the hopping matrix, which is chosen from $t_{\bf ij}\in[t-\Delta/2,t+\Delta/2]$ and zero otherwise with a probability $P(t_{\bf ij})=1/\Delta$. The strength of disorder can be characterized by $\Delta$, which represents the magnitude of the modification of matrix elements $t_{\bf ij}$. The parameter $t$ is set as $t=1$ as the energy scale. Here $U>0$ represents the on-site repulsive interaction. In this article, we mainly use $U=3|t|$. In the presence of disorder, reliable results are obtained by taking an average of 20 disorder simulations, as it has been demonstrated to effectively avoid errors introduced by randomness \cite{PhysRevLett.120.116601,PhysRevB.105.045132}.

Our simulations are mostly performed on lattices of double-48 sites with periodic boundary conditions. The double-48 lattice implies a total number of sites $N_s=2\times3\times4^2=96$ \cite{PhysRevB.84.121410}. The finite-temperature determinant quantum Monte Carlo(DQMC) method is employed to complete simulations. The basic strategy of the DQMC is to express the partition function $Z=Tr e^{-\beta H}$ as a path integral over a set of random auxiliary fields. The imaginary time interval $(0,\beta)$ is discretely divided into $M$ slices of interval $\Delta \tau $, which is chosen as small as 0.1 to control the ``Trotter errors."

To investigate the superconducting property of graphene, the pairing susceptibility is computed:
\begin{equation}
P_{\alpha}=\frac{1}{N_s}\sum_{i,j}\int_{0}^{\beta}d\tau\langle\Delta_{\alpha}^{\dag} (i,\tau)\Delta_{\alpha}(j,0)\rangle,
\label{Pa}
\end{equation}
where $\alpha$ stands for the pairing symmetry. Due to the constraint of the on-site Hubbard interaction in Eq.(\ref{hamiltonian}), pairing between two sublattices is favored and the corresponding order parameter $\Delta_{\alpha}^{\dag} (i)$ is defined as

\begin{equation}
\Delta_{\alpha}^{\dag}(i)=\sum_{l}f_{\alpha}^{\dag}(\delta_l)(c_{i\uparrow}c_{i+\delta_l \downarrow}- c_{i\downarrow}c_{i+\delta_l\uparrow})^{\dag}
\label{Da}
\end{equation}
with $f_{\alpha}(\delta_l)$ being the form factor of pairing function. The vector $\delta_l(l=1,2,3)$ denotes the nearest-neighbor(NN) connection. Considering the honeycomb lattice symmetry of the D6 point group, two possible NN pairing symmetries are characterized: (i) $ES$ wave, and(ii) $d+id$ wave.

\begin{equation}
{ ES\ \rm wave}: f_{ES}(\delta_l)=1,  \quad  l=1,2,3
\label{ESwave}
\end{equation}

\begin{equation}
{ d+id\ \rm wave}: f_{d+id}(\delta_l)=e^{i(l-1)\frac{2\pi}{3}},   \quad  l=1,2,3
\label{d+idwave}
\end{equation}

\noindent

\section{Results and discussion}

\begin{figure}[htbp]
\centering
\includegraphics[scale=0.5]{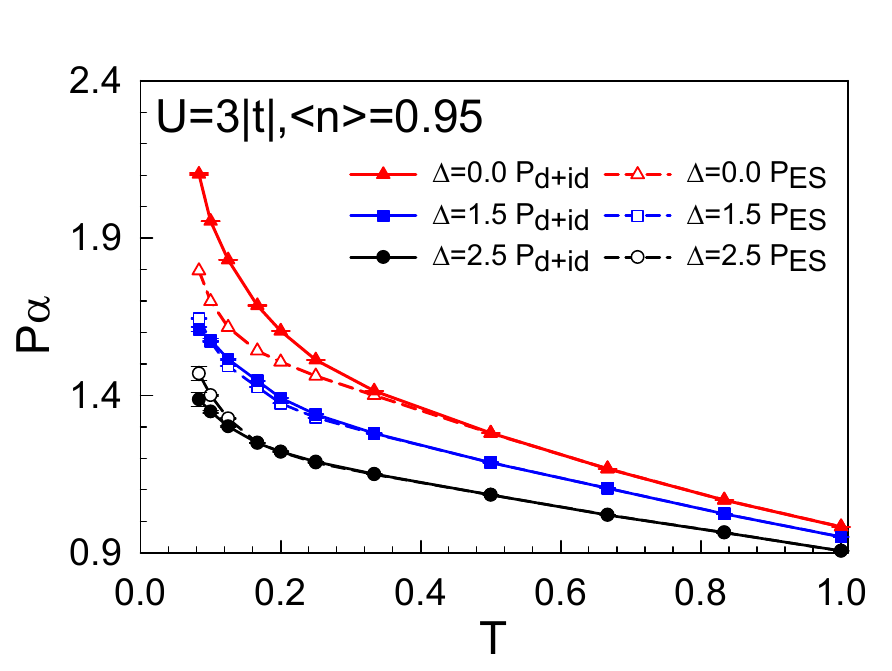}
\caption{Pairing susceptibility $P_{\alpha}$ as a function of temperature $T$ for different pairing symmetries $P_{ES}$, $P_{d+id}$, and different disorder strength $\Delta$ with electron density $\avg{n}=0.95$. }
\label{Fig2}
\end{figure}

We first investigated the variation of the pairing susceptibility $P_{ES}$ and $P_{d+id}$ with temperature for the electron density close to half filling, $\avg{n}=0.95$, as shown in Fig.\ref{Fig2}. The solid lines represent the $d+id$ pairing symmetries while the dashed lines represent the $ES$ pairing symmetries. By choosing the disorder strengths $\Delta=0.0, 1.5,$ and $2.5$, it can be observed that $P_{\alpha}$ always increases with decreasing temperature, and the bond disorder suppresses $P_{\alpha}$. In the clean limit, i.e., $\Delta=0.0$, $P_{d+id}$ increases faster than $P_{ES}$ at low temperatures, indicating the dominance of the $d+id$ wave over the $ES$ wave. At $\Delta=1.5$, the values of the two symmetries are almost equal within the range of our calculations. With a larger disorder strength, at $\Delta=2.5$, $P_{ES}$ increases faster than $P_{d+id}$ at low temperatures, indicating that the $ES$ wave dominates over the $d+id$ wave at this time. By varying the strength of the bond disorder, we find that the magnitude of $\Delta$ alters the superconducting pairing symmetry dominating.

\begin{figure}[htbp]
\centering
\includegraphics[scale=0.5]{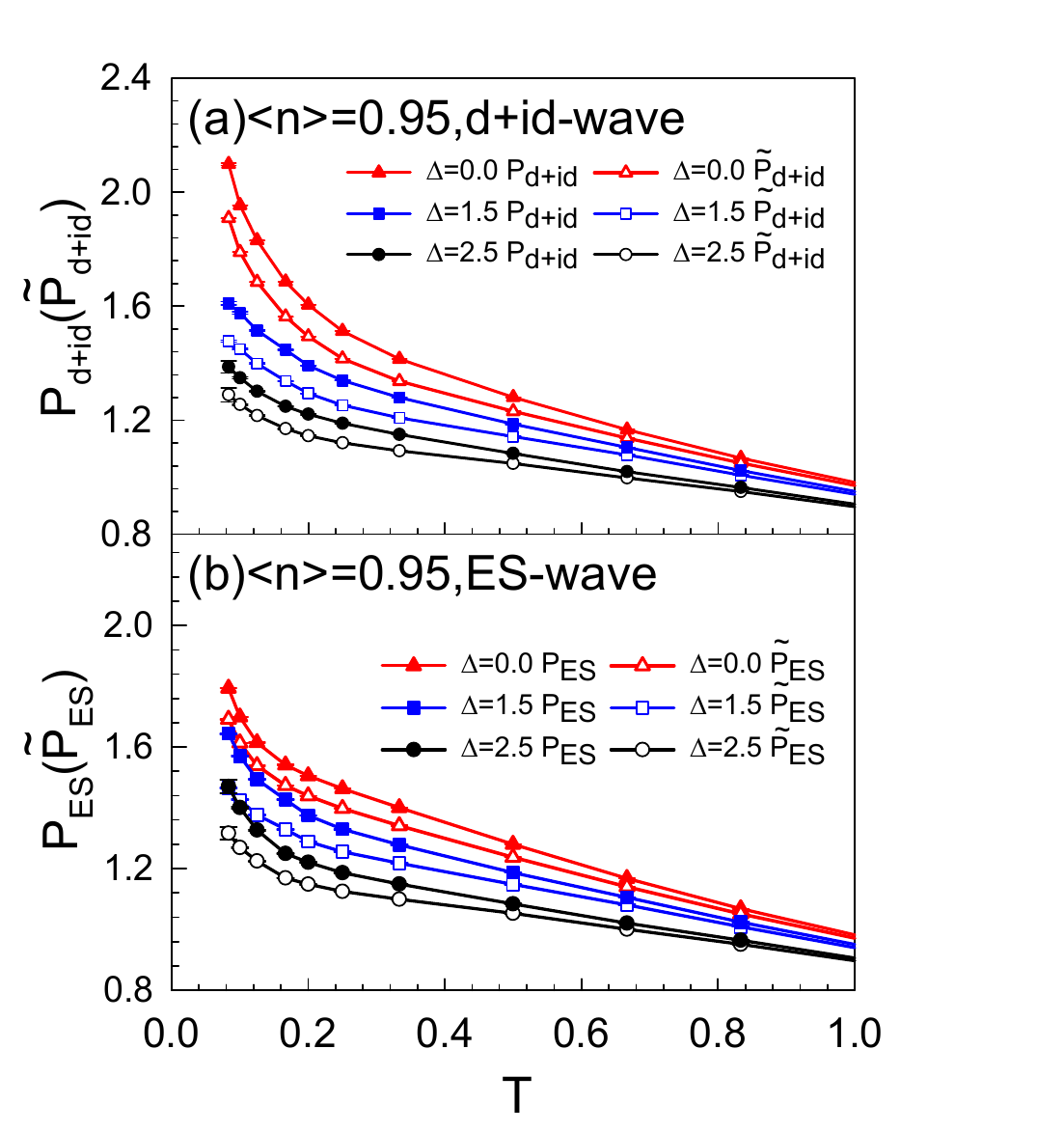}
\caption{Pairing susceptibility $P_{\alpha}$ and $\widetilde{P}_{\alpha}$ as a function of temperature $T$ for different disorder strength with electron density $\avg{n}=0.95$. (a)$d+id$ wave and (b)$ES$ wave  }
\label{Fig3}
\end{figure}

\begin{figure}[htbp]
\centering
\includegraphics[scale=0.45]{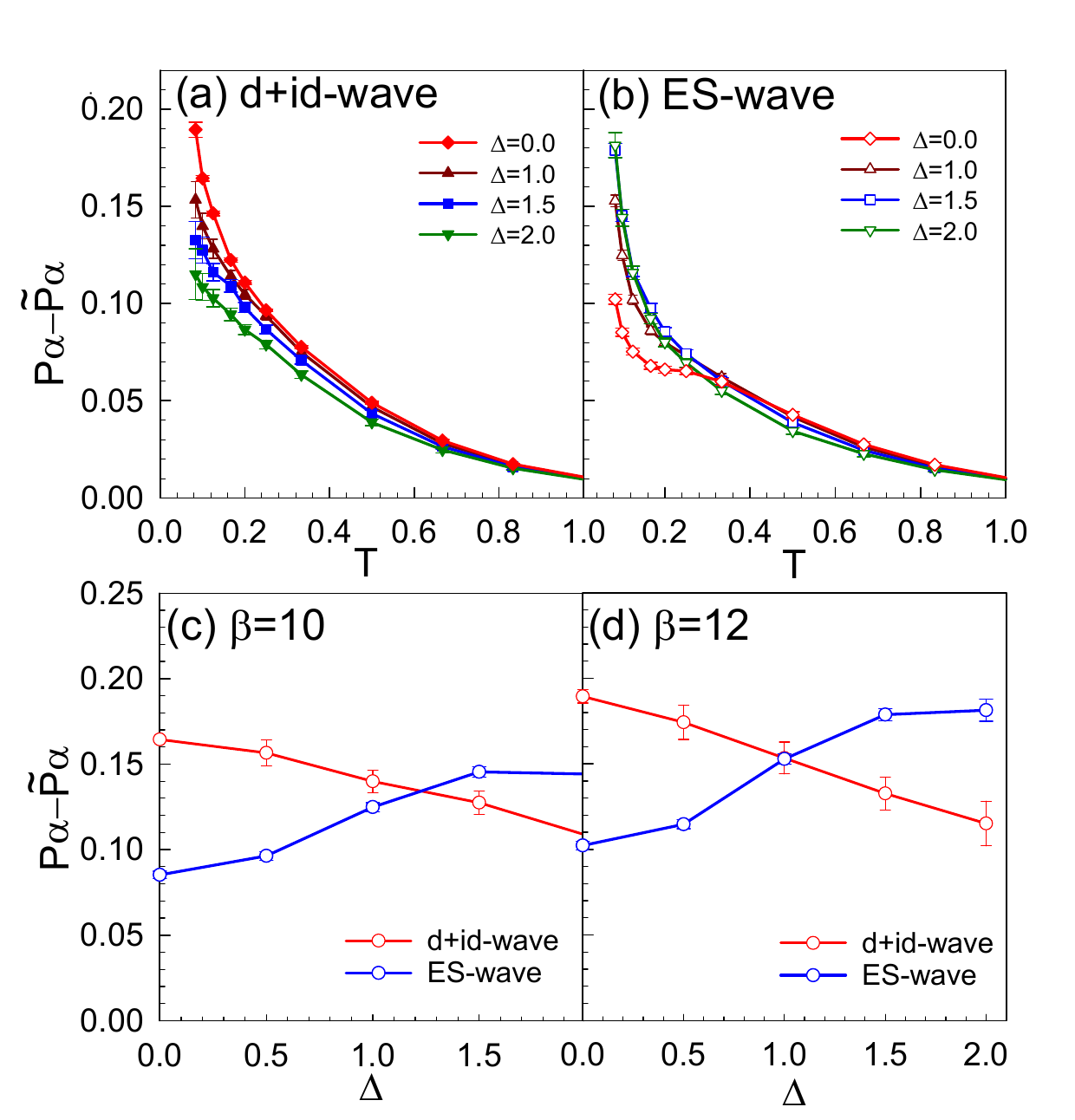}
\caption{The effective pairing interaction $P_{\alpha}-\widetilde{P}_{\alpha}$ in different pairing channels (a)$d+id$ wave and (b)$ES$ wave as a function of temperature $T$ for different disorder strength with electron density $\avg{n}=0.95$. (c) and (d) display the effective pairing interaction as a function of $\Delta$, at the fixed temperatures of (c)$\beta=10$ and (d)$\beta=12$.}
\label{Fig4}
\end{figure}

In order to make a further investigation into the superconducting pairing symmetry, we shall replace $\langle c_{i\uparrow}^{\dag}c_{j\uparrow}c_{i+l\downarrow}^{\dag}c_{j+l^{\prime}\downarrow}\rangle$  with $\langle c_{i\uparrow}^{\dag}c_{j\uparrow} \rangle\langle c_{i+l\downarrow}^{\dag}c_{j+l^{\prime}\downarrow} \rangle$ in Eq.(\ref{Pa}) to obtain the bubble contribution $\widetilde{P}_{\alpha}$, thereby extracting the effective pairing interactions in different pairing channels. In Fig. \ref{Fig3}, we compare the variations of $P_{\alpha}$ and $\widetilde{P}_{\alpha}$ with temperature for different disorder strengths $\Delta$ and pairing symmetries. Figure \ref{Fig3}(a) represents the $d+id$ wave, while Fig. \ref{Fig3}(b) represents the $ES$ wave. It can be clearly seen that with the increase in $\Delta$, both $P_{\alpha}$ and $\widetilde{P}_{\alpha}$ are suppressed, and $P_{\alpha}$ is always larger than $\widetilde{P}_{\alpha}$, which signifies that the effective pairing interactions persistently maintain positive values. To visually emphasize the influence of parameters on the effective pairing interactions, we compute $P_{\alpha}-\widetilde{P}_{\alpha}$, and the relevant results are shown in Fig. \ref{Fig4}.

Figure \ref{Fig4} shows the temperature-dependent evolution of the effective pairing correlation $P_{\alpha}-\widetilde{P}_{\alpha}$ in different pairing channels. It is evident that the effective pairing correlation is positive and tends to diverge at low temperatures in all cases, which indicates the presence of the attraction between electrons for both $ES$ and $d+id$ pairing symmetries. Figure \ref{Fig4}(a) illustrates the $d+id$ pairing symmetry. It is noteworthy that at $\avg{n}=0.95$, the effective pairing correlation $P_{d+id}-\widetilde{P}_{d+id}$ is suppressed with increasing disorder strength at low temperatures. Conversely, the effective pairing correlation $P_{ES}-\widetilde{P}_{ES}$ increases with the increasing of disorder strength at low temperatures. These results can explain the shift in dominant superconducting pairing symmetry observed in Fig.\ref{Fig2}: the introduction of disorder will suppress $P_{d+id}-\widetilde{P}_{d+id}$ while promoting $P_{ES}-\widetilde{P}_{ES}$. This implies that bond disorder may lead to a variation in the dominant pairing symmetry in graphene from $d+id$ dominant to $ES$ dominant. To illustrate the transform of the dominant pairing symmetry more clearly, we plot the effective pairing interaction as a function of $\Delta$ in Figs.\ref{Fig4}(c) and Fig.\ref{Fig4}(d) for fixed temperatures of $\beta=10$ and $\beta=12$, respectively. The results show that with increasing $\Delta$, the dominant pairing symmetry transforms from the $d+id$ to the $ES$.

\begin{figure}[htbp]
\centering
\includegraphics[scale=0.5]{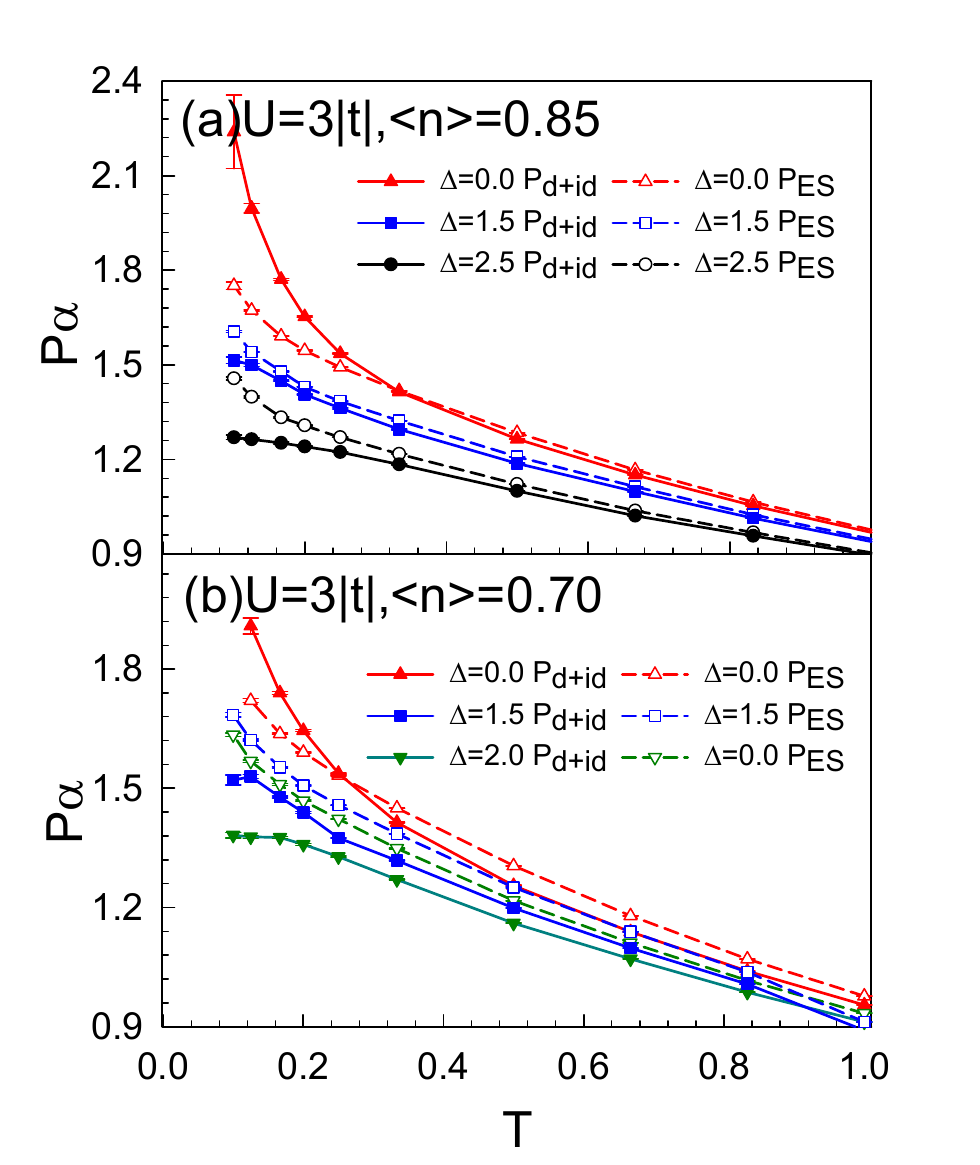}
\caption{Pairing susceptibility $P_{\alpha}$ as a function of temperature $T$ for different pairing symmetries $P_{ES}$, $P_{d+id}$, and different disorder strength $\Delta$ with electron density (a)$\avg{n}=0.85$ and (b)$\avg{n}=0.70$ }
\label{Fig5}
\end{figure}

\begin{figure}[htbp]
\centering
\includegraphics[scale=0.5]{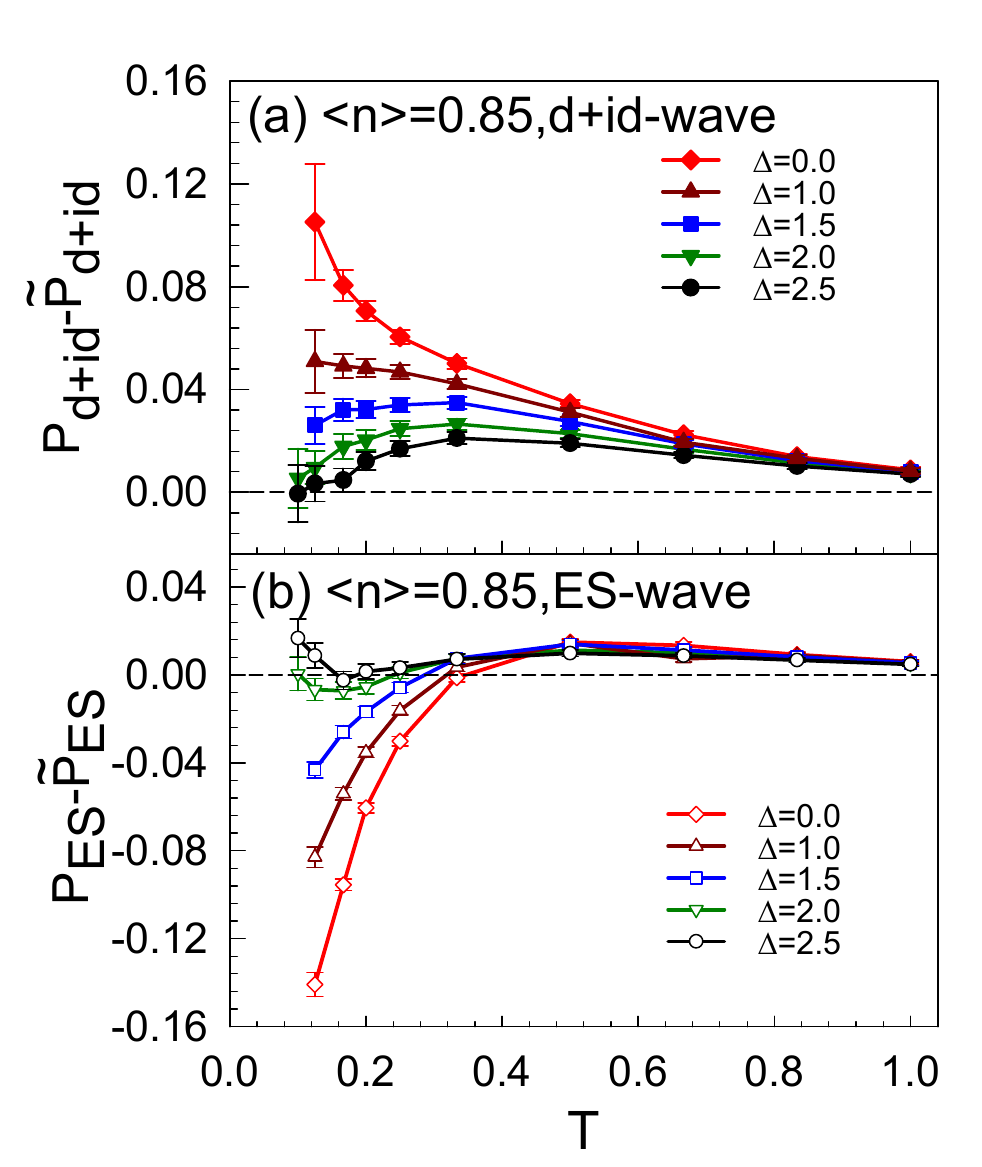}
\caption{The effective pairing interaction $P_{\alpha}-\widetilde{P}_{\alpha}$ in different pairing channels (a)$d+id$ wave and (b)$ES$ wave as a function of temperature $T$ for different disorder strength with electron density $\avg{n}=0.85$. }
\label{Fig6}
\end{figure}

\begin{figure}[htbp]
\centering
\includegraphics[scale=0.5]{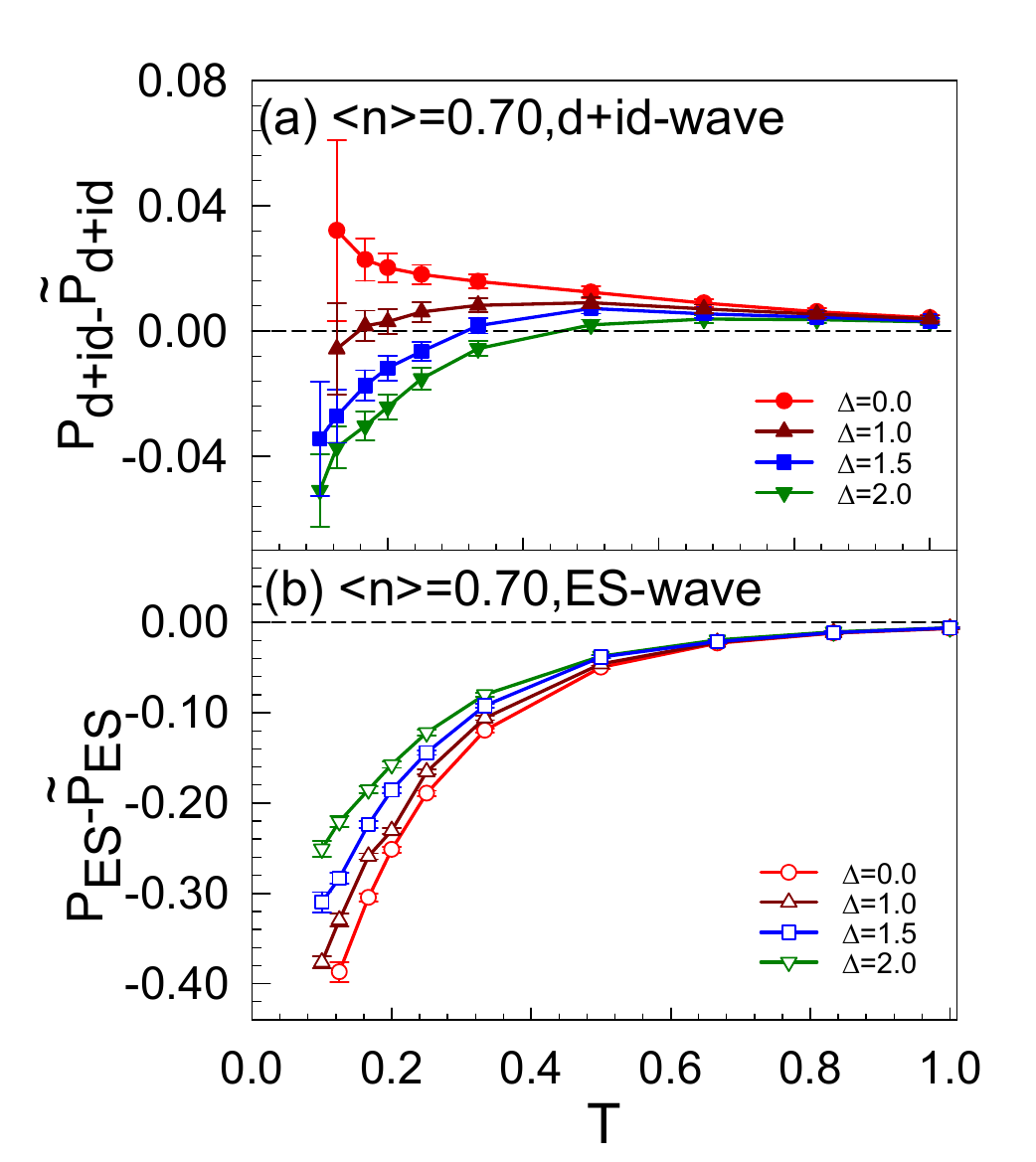}
\caption{The effective pairing interaction $P_{\alpha}-\widetilde{P}_{\alpha}$ in different pairing channels, (a) $d+id$ wave and (b) $ES$ wave, as a function of temperature $T$ for different disorder strength with electron density $\avg{n}=0.70$. }
\label{Fig7}
\end{figure}

Having investigated the situation near half-filling, we now turn our attention to regions with higher doping to see if the situation will be different. Figure \ref{Fig5} shows the temperature-dependent evolution of the pairing susceptibility $P_{ES}$ and $P_{d+id}$ at different disorder strengths for electron densities (a) $\avg{n}=0.85$ and (b) $\avg{n}=0.70$. Similar to the previous results, the solid lines represent the $d+id$ pairing symmetry and the dashed lines represent the $ES$ pairing symmetry. It can be seen that, as before, disorder shows a suppressive influence on the pairing susceptibility for both $d+id$ and $ES$ symmetries. Furthermore, with increasing disorder strength, the superconducting pairing transforms from $d+id$ wave dominance to $ES$ wave dominance. Additionally, at lower electron densities and higher disorder strength, as the temperature decreases, $P_{d+id}$ no longer diverges. This suggests the potential absence of $d+id$ wave. For a more in-depth analysis of the results, we present the temperature-dependent evolution of the effective pairing correlation $P_{\alpha}-\widetilde{P}_{\alpha}$ in different channels for $\avg{n}=0.85$ and $\avg{n}=0.70$ in Figs. \ref{Fig6} and \ref{Fig7}, respectively.

Figure \ref{Fig6} shows the temperature-dependent evolution of the effective pairing correlations (a) $P_{d+id}-\widetilde{P}_{d+id}$ and (b) $P_{ES}-\widetilde{P}_{ES}$ at $\avg{n}=0.85$. It can be found that, in the clean limit, the system exhibits positive $P_{d+id}-\widetilde{P}_{d+id}$, while negative $P_{ES}-\widetilde{P}_{ES}$ is present. The introducing of disorder, which exhibits a suppressive effect on the $d+id$ wave, leads to the disappearance of $d+id$ effective pairing correlation. In contrast, for the $ES$ wave, disorder has a promoting effect, resulting in the emergence of $ES$ wave that originally lacked $ES$ pairing. Overall, the introduction of disorder shall completely alter the superconducting pairing symmetry at $\avg{n}=0.85$. With increasing disorder strength, the system undergoes three stages: exclusively $d+id$ wave pairing, absence of both $d+id$ and $ES$ wave pairings, and exclusively $ES$ wave pairing.

As doping becomes substantial, the electron density deviates significantly from half-filling at $\avg{n}=0.70$, and the situation undergoes further changes, as shown in Fig. \ref{Fig7}. In the clean limit, only $d+id$ effective pairing exists, and the effective pairing correlation for the $d+id$ wave is not particularly large at this point. With the introduction of disorder, it is easily suppressed to become negative, indicating that there is no longer an effective attraction between electrons. Within our computational range, the effective pairing correlation for the $ES$ wave remains consistently negative, indicating that the disorder may not be sufficient to induce $ES$ pairing symmetry at this time.

\begin{figure}[htbp]
\centering
\includegraphics[scale=0.5]{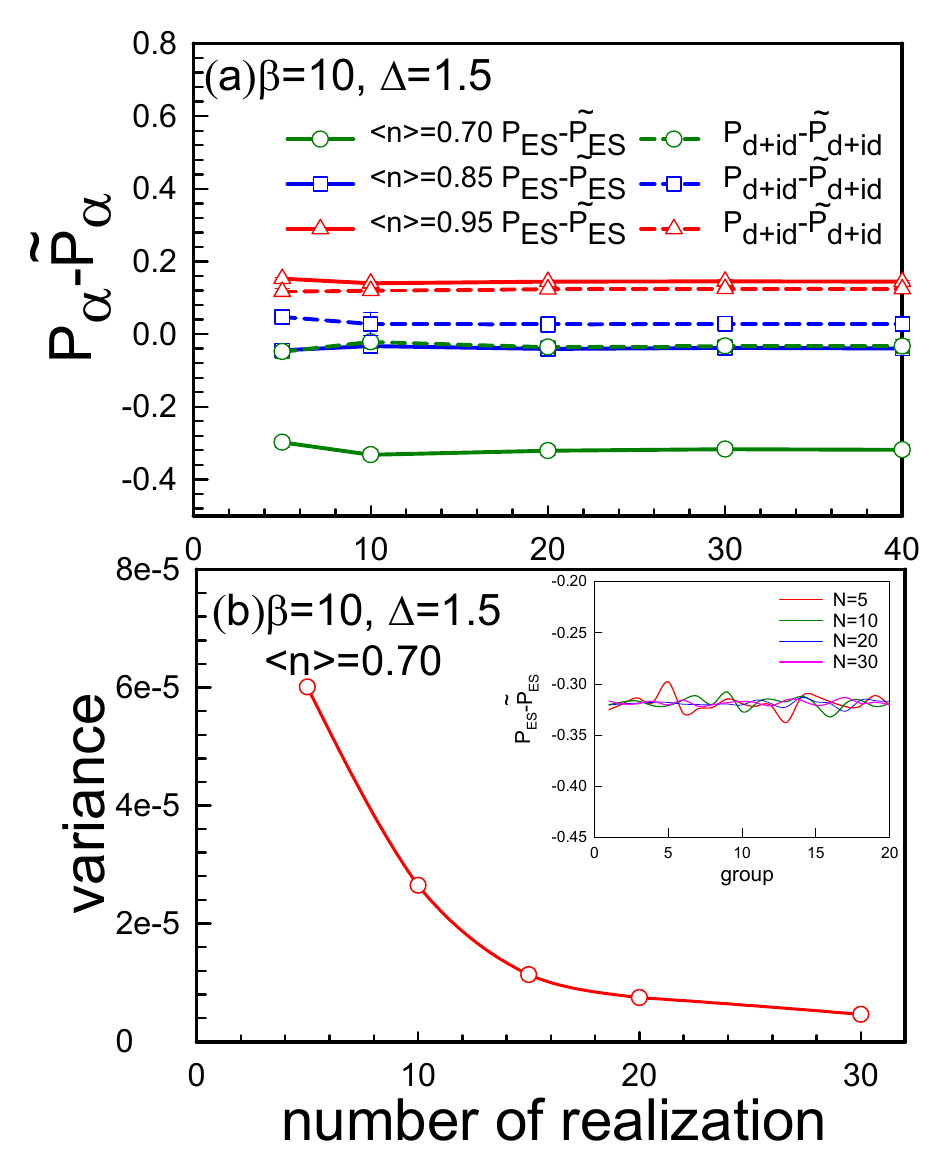}
\caption{(a) Effective pairing interaction as a function of number of realization at $\beta=10$, $\Delta=1.5$. (b) The corresponding variance of the data in the inset. Insert: The mean value of $P_{ES}-\widetilde{P}_{ES}$ as a function of the number of groups. $N$ represents the number of disorder realizations in a group.}
\label{Fig8}
\end{figure}

\begin{figure}[htbp]
\centering
\includegraphics[scale=0.45]{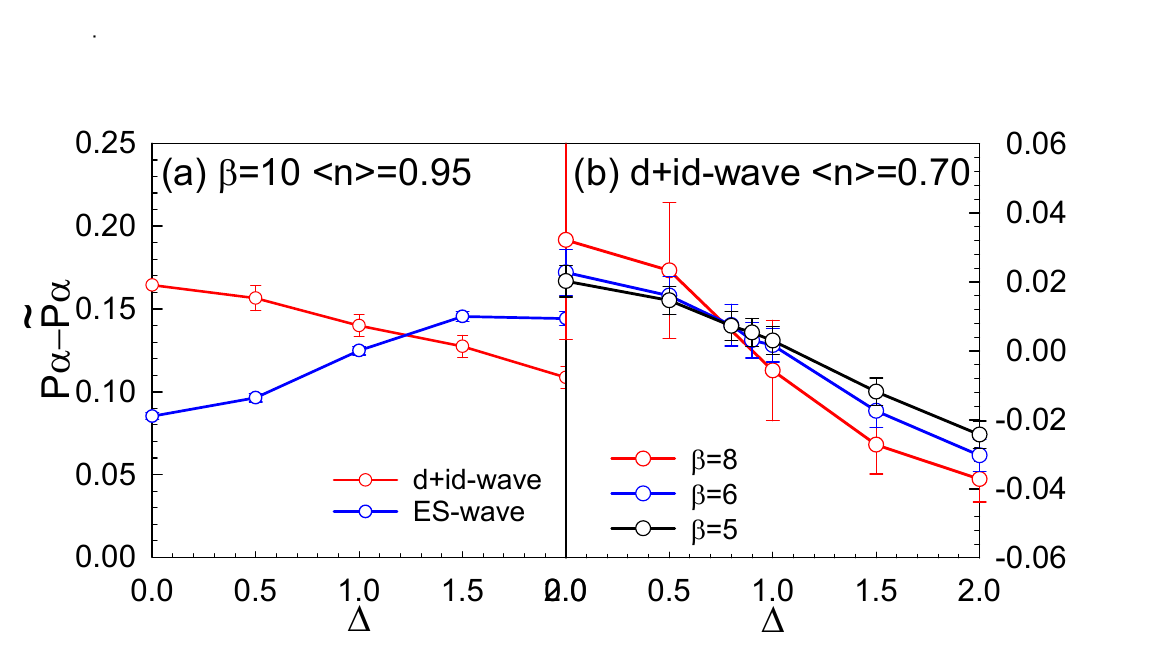}
\caption{$P_{\alpha}-\widetilde{P}_{\alpha}$ as a function of the disorder strength $\Delta$ at (a) $\beta=10$,$\avg{n}=0.95$ for different pairing symmetries and (b) $d+id$ wave at $\avg{n}=0.70$ for different temperatures.}
\label{Fig9}
\end{figure}

\begin{figure}[htbp]
\centering
\includegraphics[scale=0.45]{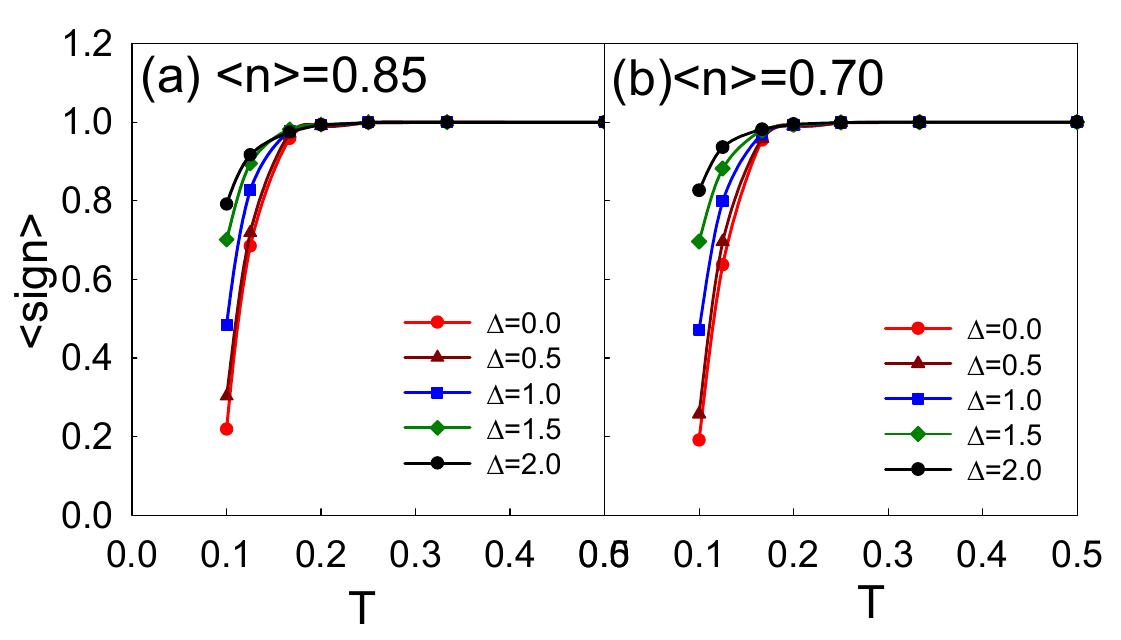}
\caption{$\avg{sign}$ as a function of the temperature for different value of bond disorder in (a) $\avg{n}=0.85$ and (b) $\avg{n}=0.70$.}
\label{Fig10}
\end{figure}

\begin{figure}[htbp]
\centering
\includegraphics[scale=0.5]{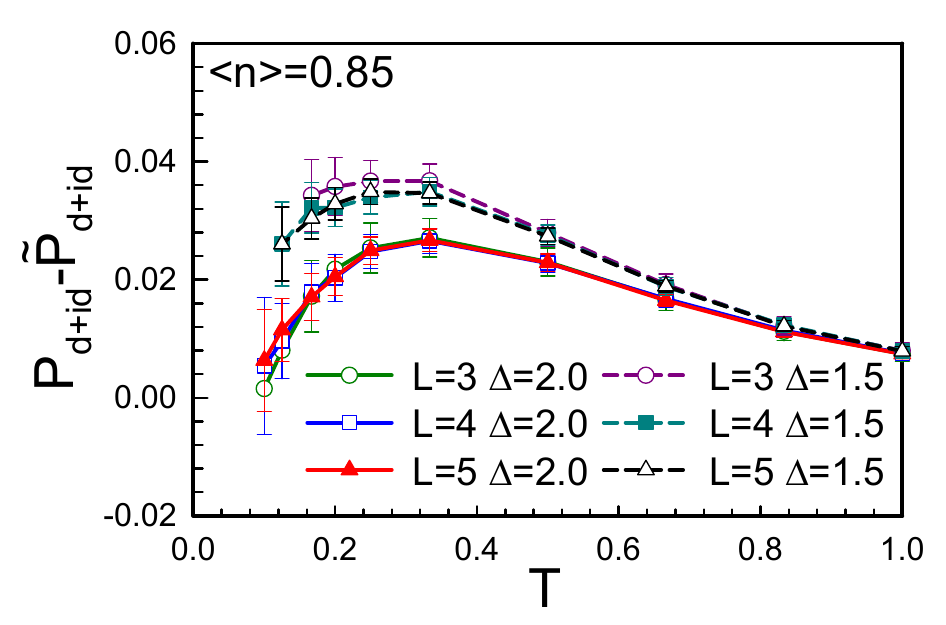}
\caption{$P_{d+id}-\widetilde{P}_{d+id}$ as a function of the temperature at $\Delta=1.5$ and $\Delta=2.0$ for $L=3,4,5$. }
\label{Fig11}
\end{figure}

\section{Conclusion}

In summary, we utilized the DQMC method to investigate the response of the superconducting order parameter to bond disorder in doped graphene. We focused on computing the pairing susceptibility $P_{\alpha}$ and effective pairing interaction $P_{\alpha}-\widetilde{P}_{\alpha}$ for three different electron densities: $\avg{n}=0.95$, $\avg{n}=0.85$, and $\avg{n}=0.70$. We observed that when the electron density was near half-filling ($\avg{n}=0.95$), both the $d+id$ wave and $ES$ wave effective pairing correlations were positive, indicating the coexistence of both pairing waves. In addition, as the disorder strength increases, the system gradually transforms from $d+id$ wave dominance to $ES$ wave dominance, suggesting that a certain strength of disorder can alter the dominant superconducting pairing. Similar behavior was observed at lower electron densities, with the distinction that for lower electron densities, both the $d+id$ and $ES$ wave effective pairing correlations were negative, indicating the absence of these two pairing symmetries. Our calculations may provide insights into the understanding of superconductivity in systems where doping and disorder coexist, as encountered in experiments.

\noindent
\section{Acknowledgements}

This work was supported by Beijing Natural Science
Foundation (No. 1242022) and Guangxi Key Laboratory of Precision Navigation Technology and Application,
Guilin University of Electronic Technology (No. DH202322). The numerical simulations in this work were performed at the HSCC of Beijing Normal University.

\appendix

\setcounter{equation}{0}
\setcounter{figure}{0}
\renewcommand{\theequation}{A\arabic{equation}}
\renewcommand{\thefigure}{A\arabic{figure}}
\renewcommand{\thesubsection}{A\arabic{subsection}}

\section{Convergence of the mean values of measurements}

The introduction of bond disorder into the system brings a certain deviation to the results due to the randomness of disorder.
To ensure the accuracy of the results, it is necessary to take the average of multiple groups of disordered results. In Fig. \ref{Fig8}(a), we show the effective pairing interaction $P_{\alpha}-\widetilde{P}_{\alpha}$ with the number of disorder realizations.
For any given density $\avg{n}$, the data does not exhibit significant changes after the number of realizations exceeds ten. We also show the variation as a function of the number of disordered realizations in Fig. \ref{Fig8}(b), and the variance curve shows good convergence. In the inset of Fig.\ref{Fig8}(b), we calculate the average of several sets of data with the number of realizations $N$ set to 5, 10, 20, and 30, respectively. The results show that the fluctuations are more severe when $N$=5, and are significantly suppressed as $N$ is further increased. This confirms the rationality of using 20 realizations in our main text.

\section{The critical point in Fig.\ref{Fig:phase}}

To determine the critical disorder strength for the transform between dominant SC pairings in Fig. \ref{Fig:phase}, we plot the effective pairing interactions $P_{\alpha}-\widetilde{P}_{\alpha}$ as functions of $\Delta$, as shown in Fig. \ref{Fig9}. In Fig. \ref{Fig9}(a), we identify the evolution of $P_{ES}-\widetilde{P}_{ES}$ and $P_{d+id}-\widetilde{P}_{d+id}$ with disorder for $\avg{n}=0.95$ and $\beta=10$, which allow us to determine the critical point for the transform of dominant SC pairings. In Fig. \ref{Fig9}(b), we plot $P_{d+id}-\widetilde{P}_{d+id}$ as a function of $\Delta$ at different temperatures for $\avg{n}=0.70$. We find that for $\Delta<\Delta_c$, $P_{d+id}-\widetilde{P}_{d+id}$ increases as the temperature is lowered, but for $\Delta>\Delta_c$, $P_{d+id}-\widetilde{P}_{d+id}$ decreases with decreasing temperature, indicating that $P_{d+id}-\widetilde{P}_{d+id}$ may become negative and is completely suppressed at low temperatures.

\section{Sign problem}

Away from half-filling usually leads to sign problems. Related researches have shown that the presence of bond disorder may alleviate the sign problem \cite{PhysRevLett.115.240402}, which is beneficial for our calculations. In Fig.\ref{Fig10}, we plot the average fermion sign $\avg{sign}$, which is the ratio of the integral of the product of up and down spin determinants to the integral of the absolute value of the product \cite{PhysRevB.92.045110}
\begin{eqnarray}
\label{sign}
\langle S \rangle &=
\frac
{\sum_{\cal X} \,\,
{\rm det} M_\uparrow({\cal X}) \,
{\rm det} M_\downarrow({\cal X})
}
{
\sum_{\cal X}  \,\,
| \, {\rm det} M_\uparrow({\cal X}) \,
{\rm det} M_\downarrow({\cal X}) \, |
}
\end{eqnarray}
as a function of temperature. When $\avg{n}=0.95$, the system is near half-filling, and the impact of the sign problem is weak, so we mainly show the average sign for (a) $\avg{n}=0.85$ and (b) $\avg{n}=0.70$. As is seen, the sign problem is particularly serious at low temperatures. Fortunately, disorder shall limit and weaken the impact of the sign problem. In order to obtain the same quality of data as $\avg{sign}\approx 1$, much longer runs are necessary to compensate the fluctuations. Indeed, we can estimate that the runs need to be stretched \cite{santos2003introduction, PhysRevD.24.2278, PhysRevB.94.075106} by a factor on the order of $\avg{sign}^{-2}$. In our simulations, especially in the simulation results where the sign problem is much worse, we have increased measurement from 10000 to 100000 times to compensate the fluctuations, and thus, the results for current parameters are reliable.

\section{Finite-size effect}

To make our diagram more convincing, we also checked results on different lattice size. In Fig.\ref{Fig11}, we take $\avg{n}=0.85$, $\Delta=1.5$, and $\Delta=2.0$ as examples, and plot $P_{d+id}-\widetilde{P}_{d+id}$ as a function of temperature for different lattice sizes $L=3$, $L=4$, and $L=5$. It can be seen that the finite size has little effect on the results, and there is no qualitative difference in the results at low temperatures, indicating that the $L=4$ we have chosen can accurately reflect the dominant pairing.

\bibliography{reference}
\end{document}